\documentclass[lettersize,journal]{IEEEtran}
\usepackage{amsmath,amsfonts}
\usepackage{algorithmic}
\usepackage{algorithm}
\usepackage{array}
\usepackage[caption=false,font=normalsize,labelfont=sf,textfont=sf]{subfig}
\usepackage{textcomp}
\usepackage{stfloats}
\usepackage{url}
\usepackage{verbatim}
\usepackage{graphicx}
\usepackage{cite}
\usepackage{multirow}
\usepackage{pifont}
\usepackage{orcidlink}

\begin{document}

\title{Diverse Instance Generation via Diffusion Models for Enhanced Few-Shot Object Detection in Remote Sensing Images}

\author{Yanxing Liu\orcidlink{0009-0007-8604-933X}, Jiancheng Pan, Jianwei Yang\orcidlink{0000-0001-5228-8462}, Tiancheng Chen\orcidlink{0009-0005-4847-4429}, Peiling Zhou and Bingchen Zhang
\thanks{Yanxing Liu, Tiancheng Chen and Bingchen Zhang are with the National Key Laboratory of Microwave Imaging Technology, Aerospace Information Research Institute, Chinese Academy of Sciences, Beijing 100094, China, and also with the School of Electronic,
Electrical and Communication Engineering, University of Chinese Academy of Sciences, Beijing 100049, China. (email: liuyanxing21@mails.ucas.ac.cn;  chentiancheng21@mails.ucas.ac.cn;  zhangbc@aircas.ac.cn)}
\thanks{Jiancheng Pan is with the department of Earth System Science, Tsinghua University, Beijing 100084, China.(email: jiancheng.pan.pluse@gamil.com)}
\thanks{Jianwei Yang and Peiling Zhou are with the Aerospace Information Research Institute, Chinese Academy of Sciences, Beijing 100094, China, also with the Key Laboratory of Technology in Geo-Spatial Information Processing and Application System, Chinese Academy of Sciences, Beijing 100190, China, and also with the School of Electronic, Electrical and Communication Engineering, University of Chinese Academy of Sciences, Beijing 101408, China.(emial:yangjianwei20@mails.ucas.ac.cn; zhoupeiling21@mails.ucas.ac.cn)}}



\maketitle

\begin{abstract}
Few-shot object detection (FSOD) aims to detect novel instances with only a limited number of labeled training samples, presenting a challenge that is particularly prominent in numerous remote sensing applications such as endangered species monitoring and disaster assessment.
Existing FSOD methods for remote sensing images (RSIs) have achieved promising progress but remain constrained by the limited diversity of instances. 
To address this issue, we propose a novel framework that can leverage a diffusion model pretrained on large-scale natural images to synthesize diverse remote sensing instances, thereby improving the performance of few-shot object detectors.
Instead of directly synthesizing complete remote sensing images, we first generate instance-level slices via a specialized slice-to-slice module, and then embed these slices into full-scale imagery for enhanced data augmentation.
To further adapt diffusion models for remote sensing scenarios, we develop a class-agnostic image inversion module that can invert remote sensing instance slices into semantic space.
Additionally, we introduce contrastive loss to semantically align the synthesized images with their corresponding classes.
Experimental results show that our method has achieved an average performance improvement of 4.4\% across multiple datasets and various approaches. Ablation experiments indicate that the elaborately designed inversion module can effectively enhance the performance of FSOD methods, and the semantic contrastive loss can further boost the performance.

\end{abstract}

\begin{IEEEkeywords}
Diffusion model, few-shot object detection, optical remote sensing imagery.
\end{IEEEkeywords}

\section{Introduction}
\IEEEPARstart{I}{n} the past decade, visual object detection~\cite{fu2025ntire,pan2025enhance,li2025semantic,li2025domain} has made significant progress, thanks to the remarkable power of deep-learning methods.
Nonetheless, most current detectors heavily rely on large-scale training samples to achieve satisfactory performance, while in practice, data acquisition and annotation within the field of remote sensing are both challenging~\cite{pan2025earthsynth}. 

Obtaining sufficient labeled data is time-consuming and, for many applications, may even be infeasible. 
When only a few training samples are available, detectors based on deep learning could suffer from overfitting and low detection performance. \par{}
To address these limitations, several works~\cite{li2021few, liu2024few} have been proposed to study more effective few-shot detectors in remote sensing images (RSIs). 
Concretely, most FSOD approaches can be roughly divided into fine-tuning~\cite{liu2024few,chen2022multiscale} methods and meta-learning~\cite{li2021few} methods.
For example, Li et al.~\cite{li2021few} propose Prototype-CNN, which leverages a prototype learning network (PLN) and a prototype-guided RPN (P-G RPN) to detect few-shot novel instances. 
MSOCL~\cite{chen2022multiscale} incorporates multiscale object contrastive
learning to more fully represent object features by adopting a Siamese network structure in few-shot training. 
Liu et al.~\cite{liu2024few} propose a label-consistent classifier and a coarse-to-fine RPN to alleviate class and location biases. 
MP-FSDET~\cite{liu2024multi} further employs multi-modal prototypes to compensate for the limitations of single-modal features, while LAE-DINO~\cite{pan2025locate} applies multi-modal features to open-set detection tasks.	
\par{}
However, although previous approaches have demonstrated their effectiveness, they only trained detectors on quite a few labeled samples and do not increase the diversity of instances. 
Currently, large-scale pretrained diffusion models have made tremendous progress in the computer vision field, and they can generate diverse instances by leveraging large-scale pretrained samples.
Zhang et al.~\cite{zhang2025controllable} have explored the application of condition diffusion models for direct image generation to improve the performance of FSOD in RSIs.
Control Copy-Paste~\cite{liu2025control} proposed a controllable diffusion-based method to enhance the performance of FSOD by leveraging diverse contextual information.	
However, we observed that this approach may generate noisy objects beyond the conditioned regions.
Inspired by this, we propose Diverse Instance Generation based Few-Shot Object Detection for Remote Sensing Images (DIG-FSOD), a framework that aims to leverage a diffusion model to generate diverse object instances, thereby enhancing the performance of few-shot object detection in RSIs. 
To avoid the generation of unexpected noise targets, we do not directly generate the complete remote sensing images. 
Instead, we first generate slices of remote sensing instances and subsequently embed these slices into a full image. \par{}
Specifically, we first propose the Hybrid Image Inversion Module (HIIM) that inverts the remote sensing instances to the unified latent space to bridge the gap between existing diffusion models based on natural images and those designed for remote sensing slices.
In contrast to the previous approach~\cite{zhang2024ssr} that only use CLIP to invert images to the textual latent space, we simultaneously employ both detail-rich self-supervised features~\cite{oquab2023dinov2} and semantically-rich CLIP features.
In addition, to ensure the model focuses more on the semantic features of target instances rather than simply repeating reference instances, we propose a semantic contrastive loss~\cite{pan2023reducing,pan2023prior,10507076}. 
To enable the current diffusion model to generate high-quality remote sensing instance slices, we employ a two-stage training strategy to fully leverage existing remote sensing datasets. 
In the first stage, we train a class-agnostic HIIM using data of base classes. This module can invert few-shot support images into the textual condition space and use the few-shot images as reference images to generate target instances.
In the second stage, HIIM is fine-tuned on few-shot novel instances, thereby enhancing semantic features through class contrastive learning.
Extensive comparative and ablation experiments conducted on DIOR~\cite{li2020object} and NWPU VHR-10~\cite{cheng2014multi} datasets demonstrate that our method effectively improves detector performance in remote sensing few-shot object detection. \par{}
In summary, the main contribution of this article can be summarized as follows.\par{}
1. We propose an efficient FSOD framework based on diffusion model, which enables existing diffusion models to generate high-quality RSIs by inverting condition RSIs to domain-agnostic semantic spaces, and enhances semantic features through additional semantic contrastive loss. \par{}
2. We design the HIIM that simultaneously leverages detail-rich self-supervised features and semantically rich CLIP features to generate high-quality remote sensing images.\par{}
3. To further boost FSOD performance and ensure semantic consistency in generated images, we apply a contrastive loss between global features of condition image and semantic textual features.

\section{Method}
\subsection{Overall Framework}
\begin{figure*}[t]
\centering
\includegraphics[width=7in]{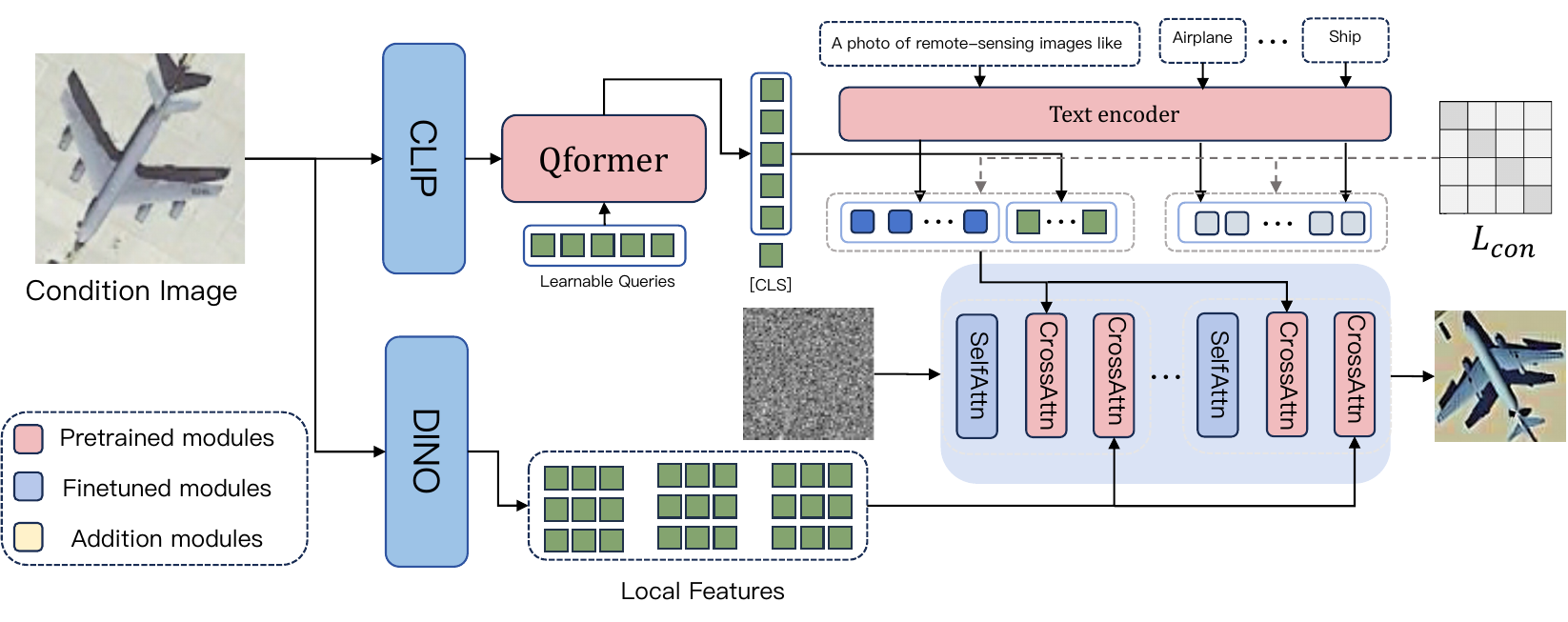}
\caption{Overall framework of our proposed DIG-FSOD. To keep things simple, the VAE encoder-decoder architecture is not shown in the illustration. Our method first encodes the condition image into global and local latent features through dedicated encoders, and then leverages a conditional diffusion model in the latent space to generate new latent features. The generated instances are finally generated into large remote sensing images using the AnyDoor. During training, global-text alignment is reinforced using a contrastive loss, while the final reconstructed images are supervised by both mean-square error and contrastive losses.}
\label{fig_2}
\end{figure*}

The overall framework of our method is based on the latent diffusion model~\cite{rombach2022high} and is illustrated in Fig.~\ref{fig_2}.
First, the global concept encoder and local feature encoder $\mathcal{E}_g(\cdot),\mathcal{E}_{l}(\cdot)$ are trained to invert a condition remote sensing image $x$ to lower-dimension latent spaces. 
The process can be formulated as $f_g,f_l = \mathcal{E}_g(T(x)),\mathcal{E}_l(T(x))$, where $T(\cdot)$ denotes some basic augmentations like flipping and rotation to avoid simply copying the original image.
To reduce computational cost, $x$ will be mapped to a lower-dimensional latent space $z = \mathcal{E}(x)$ by an encoder trained from VAE. 
Then, the condition diffusion model $\epsilon_{\theta}(\cdot)$ will be trained on the latent space to generate latent codes based on both global and local conditions $f_g$ and $f_l$. 
We first adopt the mean-squared loss (MSE) to train the diffusion model:
\begin{equation}
    L_{ldm} := \mathbb{E}_{z, f_g, f_l, \epsilon \sim \mathcal{N}(0,1), t} \left[ \left\| \epsilon - \epsilon_\theta (z_t, t, f_g, f_l)\right\|_2^2 \right]
\end{equation}
where $\epsilon$ denotes the unscaled gaussian noise, $t$ is the time step, $z_t$ is the latent noise at time step $t$, $f_g$ and $f_l$ represent the global and local features extracted from $\mathcal{E}_g(\cdot)$ and $\mathcal{E}_{l}(\cdot)$.
In the inference phase, a random gaussian noise $z_T$ is iteratively denoised to $z_0$ and decoded into the pixel space by $x_{\prime} = \mathcal{D}(z_0)$, where $\mathcal{D}$ denotes the decoder trained in VAE.
In addition to the original MSE loss, we also employ a contrastive loss to maintain consistency between global features and textual semantic features. \par{}
The details description of our proposed HIIM will be introduced in section \ref{sec:2.2}.
The generated image is supervised by the reconstruction loss of MSE and an additional contrastive learning loss in section \ref{sec:2.3}.

\subsection{Hybrid Image Inversion Module}\label{sec:2.2}
The Hybrid Image Inversion Module (HIIM) contains a global concept encoder, which aims to invert the condition image into a textual latent space, and a local feature encoder, which aims to maintain the detailed info of the condition image. 
Unlike text-to-image (T2I) tasks, image-to-image (I2I) tasks demand not only the ability to generate images but also a comprehensive understanding of the condition image. 
The understanding of condition image requires semantic information from the condition image, whereas image generation tasks need more fine-grained features from the condition image.
Therefore, we use CLIP in the global concept encoder to extract semantic information, while employing DINOv2~\cite{oquab2023dinov2} in the local feature extractor to obtain more fine-grained details. \par{}
Local features serve as extra conditional information in the generation process of the diffusion model through additional cross-attention layers.
We incorporate global features in the same manner as the original Stable Diffusion and utilize an additional cross-attention layer to integrate local features into the query process. 

\subsubsection{Global concept encoder}
The clip features are used in the global concept encoder to provide global semantic features. 
The [CLS] feature in CLIP maintains rich semantic information about the target and is aligned with the text encoder, so we directly use it as a semantic feature.
However, a single [CLS] feature is not sufficient for a comprehensive understanding of the condition image.
Therefore, we use Qformer to provide additional semantic information about the target.
Here, the Qformer denotes a transformer module where the queries are learnable queries, and the clip features are input as a part of keys and values.
We also use PerceiverAttention to aggregate queries and clip features. The detailed calculation is shown in
\begin{equation}
    f_{g}^{\prime} = Softmax(\frac{Q(f_q)K([f_{clip},f_q])}{\sqrt{d}})V([f_{clip},f_q]) \label{eq_1}
\end{equation}
where the $f_q$ denotes learnable queries and $f_{clip}$ denotes the features of clip.\par{}
At last, the [CLS] feature of clip and $f_g^\prime$ will be concatenated into the global feature $f_g$, as described in Eq. \ref{eq_2}.
\begin{equation}
    f_g = Concat(f_{[CLS]},f_g^{\prime}) \label{eq_2}
\end{equation}

\subsubsection{Local feature encoder}
Relying solely on global features can hinder the model from focusing on the discriminative local features of the condition image, thereby diminishing the reliability of the generated image. 
Therefore, we employ the multi-level patch features derived from DINOv2~\cite{oquab2023dinov2} to provide fine-grained references for the generation process, as self-supervised models contain richer object details compared to models using cross-modal contrastive learning. \par{}
As demonstrated in Janus, self-supervised features and text-vision contrastive features each exhibit distinct advantages and disadvantages.
Therefore, we employ multi-layer DINOv2~\cite{oquab2023dinov2} features to provide discriminative local features for the reference image. 
Specifically, the local feature extractor processes a condition image to produce multi-scale detailed image features $z_I=\{z_k\}_{k=0}^{K}$, where $z_k$ represents the visual features at the scale of k in DINOv2 visual backbone and $K$ refers to the number of target scales. We set K to 5 in all experimental settings. \par{}
The detailed calculation of local feature extractor is demonstrated in Eq. \ref{eq_3}.
\begin{equation}
    f_l = Concat(\{z_k\}_{k=0}^K) \label{eq_3}
\end{equation}
The local features are conditioned into the reverse diffusion process through additional cross attention layers.

\subsection{Class Contrastive Learning}\label{sec:2.3}
To enhance the semantic information of the target image as much as possible, we employ contrastive learning to supervise the generative model.
Nevertheless, performing contrastive learning directly on the generated images and class condition in a manner similar to CLIP necessitates decoding the corresponding latent representations during the training phase.
Therefore, we project the global feature into the textual semantic space via a single-layer perceptron layer, subsequently aligning it with the textual representations using the InfoNCE loss function. 
We use the InfoNCE loss between the generated images and the text embeddings from CLIP, in order to ensure that the generated images preserve as much of the corresponding target’s semantic information as possible.
The loss function can be formulated as 
\begin{equation}
    L_{con} = -\frac{1}{N}\sum_{i=1}^{N}log\frac{exp(\frac{MLP(f_g )\cdot f_t^{+}}{\tau})}{\sum_{j=1}^{N}exp(\frac{MLP(f_g)\cdot f_t^{-}}{\tau})}
\end{equation}
where N is batch size, $MLP(\cdot)$ denotes the Multilayer Perceptron projection function used in Qformer and $f_t$ represents the CLIP feature associated with the classes. \par{}
The contrastive loss function effectively improves the consistency between generated instances and semantic classes, thereby ensuring that the generated instances accurately capture the semantic information inherent in the target classes.
\section{Experiments}
\subsection{Datasets, Evaluation and Implementation Details}
\subsubsection{Datasets} 
To verify the effectiveness of our proposed method, we conduct experiments on DIOR~\cite{li2020object} and NWPU VHR-10~\cite{cheng2014multi} datasets.\par{}
DIOR~\cite{li2020object} is a large-scale public remote sensing object detection dataset that contains 23463 images with 192472 objects involving 20 common categories. 
The entire dataset is divided into three parts: training set, validation set, and testing set, containing 5682, 5863, and 11738 images, respectively. 
Following TEMO~\cite{lu2023few}, we use airplane, baseball field, expressway toll station, tennis court and windmill as few-shot novel classes while the rest classes of DIOR~\cite{li2020object} dataset as set as base classes. \par{}
NWPU VHR-10~\cite{cheng2014multi} is a very high-resolution optical remote-sensing dataset. The dataset contains 800 optical images, which are composed of 150 negative samples without any annotated objects and 650 positive samples with at least one annotated object. The airplane, baseball-diamond and tennis court are set as few-shot novel classes while the rest classes of the dataset are regarded as base datasets. \par{}
We adopt the evaluation metrics used in SAE-FSDet\cite{liu2024few}, calculating mean Average Precision (mAP) for few-shot novel classes across each dataset.
\subsubsection{Implementation Details}
Most of our experiments are based on the SAE-FSDet~\cite{liu2024few}, unless otherwise specified. 
Owing to the flexible design of the data generation method, our method can be compatible with other FSOD detectors.
We use the stable diffusion v1.5, which is pretrained on LAION-5B, and use the clip-vit-g as the text encoder to enhance the text feature extraction capability of remote sensing categories.
The total training is divided into two phases. The first phase pretrains the generator on a combined dataset of DIOR~\cite{li2020object}, DOTA and FAIR1M datasets to get a class-agnostic HIIM. 
Note that to avoid class leak, we filter out all novel instances in the combined dataset.
The model is then fine-tuned on the few-shot dataset to enhance its ability to generate novel classes. 
The learning rate of $5e-5$ is set for both base training and fine-tuning stages, and we use the Adam optimizer to optimize the training parameters. 
The batch size is set to 4, the epoch of training is set to 12, and the experiments are carried out with PyTorch on an Nvidia RTX 4090 GPU.

\subsection{Experimental Results and Comparisons}
In this section, we evaluate the effectiveness of our proposed method with different FSOD methods on DIOR~\cite{li2020object} and NWPU VHR-10~\cite{cheng2014multi} datasets.

\subsubsection{Results on DIOR Dataset}
Under the challenging large-scale DIOR~\cite{li2020object} dataset,we exhibit the performance gains achieved by employing our approach with FSCE~\cite{sun2021fsce}, MSOCL~\cite{chen2022multiscale} and SAE-FSDet~\cite{liu2024few} on 3-, 5-, 10- and 20-shot settings. 
We have also adopted current remote sensing foundation diffusion model~\cite{10988859} to generate RSIs for data augmentation.
Some approaches~\cite{sun2021fsce} are designed for natural scenes and other approaches~\cite{chen2022multiscale,liu2024few} are designed for remote sensing imagery.
The detail experimental results are reported in Table \ref{tab:dior_comparation}. \par{}
\begin{table}[htbp]
    \centering
    \caption{FSOD performance on the DIOR dataset under the 3-, 5-, 10-, and 20-shot settings}
    \label{tab:dior_comparation}
    \renewcommand{\arraystretch}{1.3} 
    \begin{tabular}{ccccc}
        \hline
        Method & 3-shot & 5-shot & 10-shot & 20-shot                        \\
        \hline
        FSCE~\cite{sun2021fsce}       & 18.30 & 18.59 & 26.16 & 29.60         \\
        FSCE + Ours & 21.32 & 26.98 & 28.87 & 33.49 \\ \hline
        MSOCL~\cite{chen2022multiscale} & 25.75  & 26.16  & 32.65    & 37.36   \\
        MSOCL + Ours  & 31.05  & 33.36  &  34.52   & 42.18  \\ \hline
        SAE-FSDet~\cite{liu2024few} & 15.89 & 17.04 & 30.20 & 37.34             \\ 
        SAE-FSDet + T2E~\cite{10988859} & 16.34 & 17.20 & 29.99 & 35.97\\
        SAE-FSDet + T2E-Inpating~\cite{10988859} & 19.51  & 19.82 & 32.10 & 38.79  \\
        SAE-FSDet + Ours & 22.09 & 29.85 & 36.66 & 40.81 \\
        \hline
    \end{tabular}
\end{table}
As the table demonstrates, our method can enhance the performance of various FSOD approaches, especially when the training data is extremely limited.
As T2E models lack the ability to generate instance-level remote sensing objects, and inpainting models exhibit limited positional control in remote sensing images, the performance improvement of these methods are limited.
When employing SAE-FSDet~\cite{liu2024few} as the baseline model, our method can increase the mAP of novel classes by an average of 7.23\% across four few-shot settings, with a maximum performance improvement of 12.81\%. \par
\begin{figure*}[htbp]
  \centering
  \begin{minipage}{0.18\textwidth}
    \includegraphics[width=\textwidth]{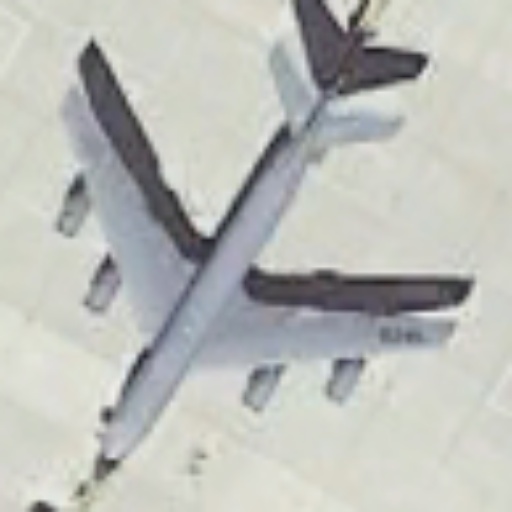}
  \end{minipage}
  \hfill
  \begin{minipage}{0.18\textwidth}
    \includegraphics[width=\textwidth]{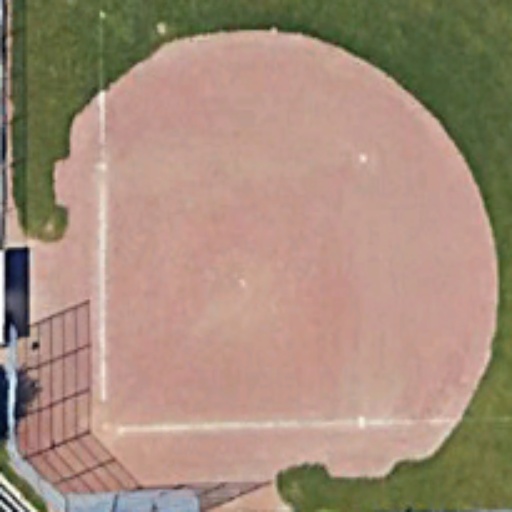}
  \end{minipage}
  \hfill
  \begin{minipage}{0.18\textwidth}
    \includegraphics[width=\textwidth]{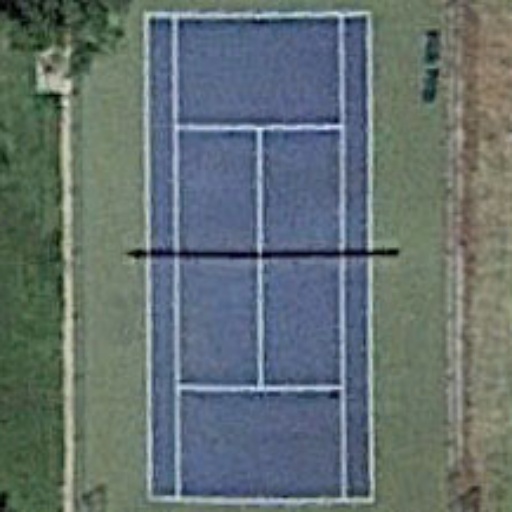}
  \end{minipage}
  \hfill
  \begin{minipage}{0.18\textwidth}
    \includegraphics[width=\textwidth]{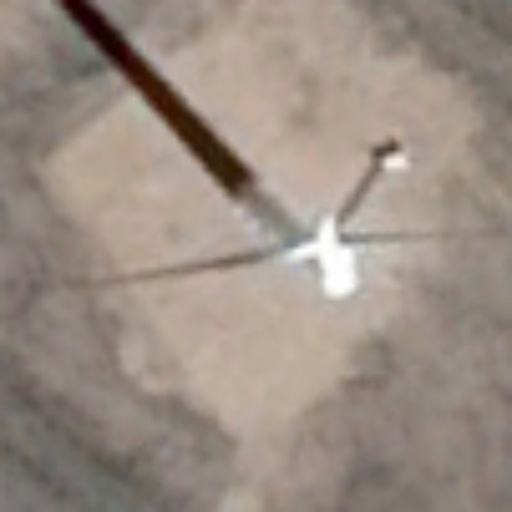}
  \end{minipage}
  \begin{minipage}{0.18\textwidth}
    \includegraphics[width=\textwidth]{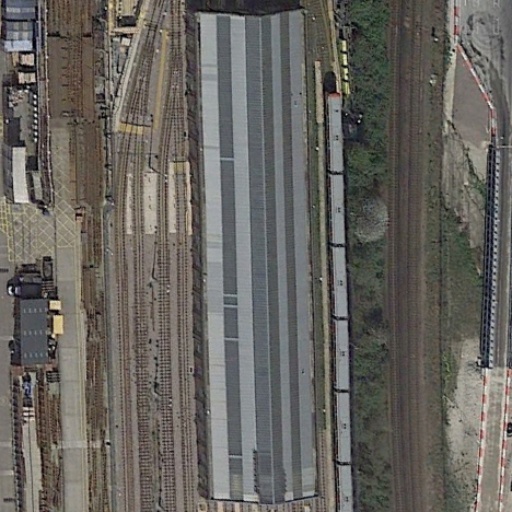}
  \end{minipage}

  \bigskip
  
  \begin{minipage}{0.18\textwidth}
    \includegraphics[width=\textwidth]{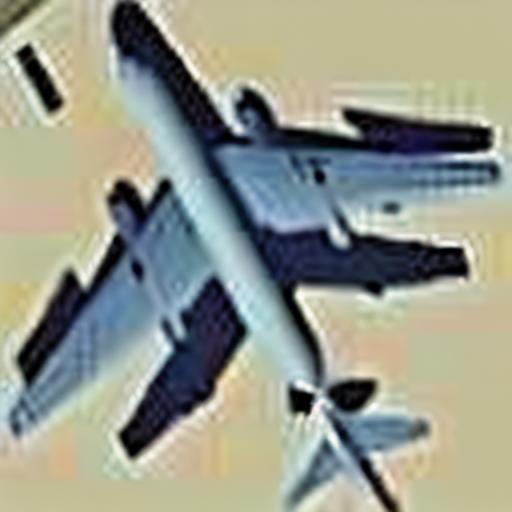}
  \end{minipage}
  \hfill
  \begin{minipage}{0.18\textwidth}
    \includegraphics[width=\textwidth]{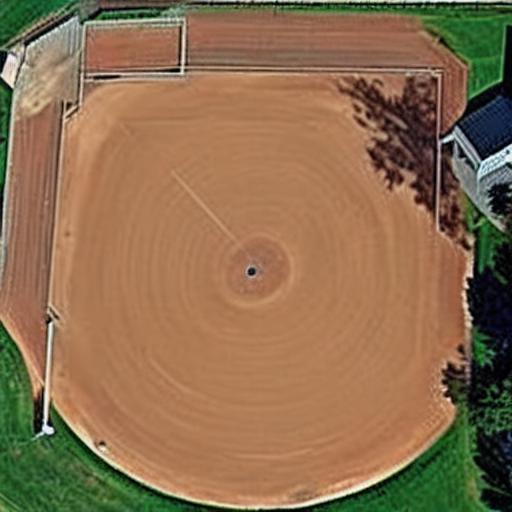}
  \end{minipage}
  \hfill
  \begin{minipage}{0.18\textwidth}
    \includegraphics[width=\textwidth]{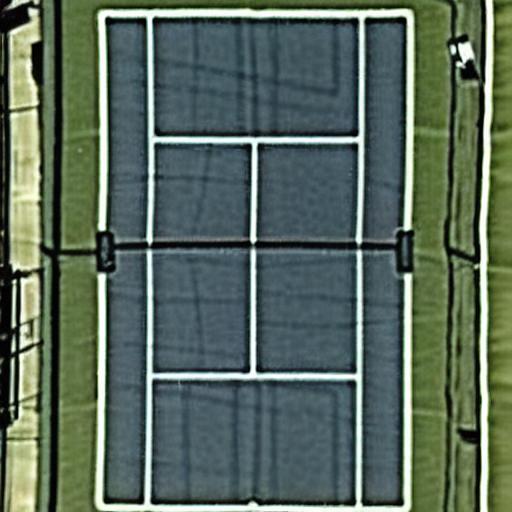}
  \end{minipage}
  \hfill
  \begin{minipage}{0.18\textwidth}
    \includegraphics[width=\textwidth]{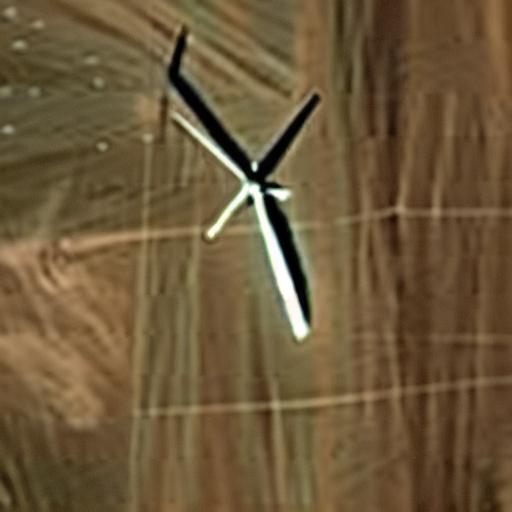}
  \end{minipage}
  \begin{minipage}{0.18\textwidth}
    \includegraphics[width=\textwidth]{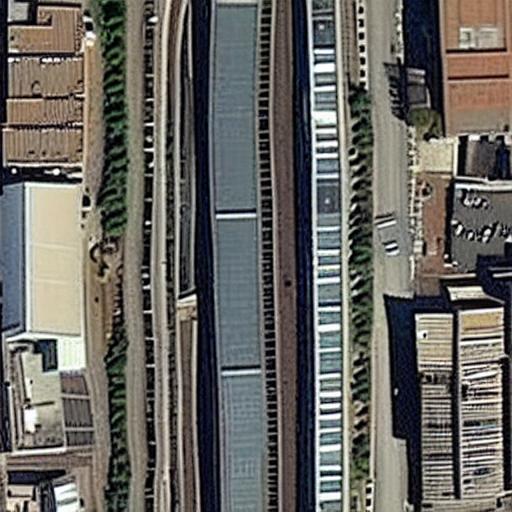}
  \end{minipage}
  \caption{Visualization of generated object instances. Instances in the first row are reference images and those in the second row are generated images.}
  \label{fig:fig3}
\end{figure*}

{}
Fig.~\ref{fig:fig3} shows some generated instances and corresponding reference images. The first row shows the reference images and the second row shows the generated image. 
As shown in Fig.~\ref{fig:fig3}, our method can increase the diversity of few-shot target instances while preserving their discriminative features.

\subsubsection{Results on NWPU VHR-10 Dataset}
To further demonstrate the effectiveness of our proposed method, we also conduct experiments on the NWPU VHR-10~\cite{cheng2014multi} dataset. We compare our method with the approach for natural scenes~\cite{sun2021fsce} and those for remote sensing images~\cite{chen2022multiscale,liu2024few}. The comparison results are reported in Table \ref{tab:nwpu_comparation}. \par{}

\begin{table}[htbp]
    \centering
    \caption{FSOD Performance on the NWPU VHR-10 dataset under the 3, 5, 10 and 20-shot setting}
    \label{tab:nwpu_comparation}
    \renewcommand{\arraystretch}{1.3} 
    \begin{tabular}{ccccc}
        \hline
        Method & 3-shot & 5-shot & 10-shot & 20-shot \\
        \hline
        FSCE~\cite{sun2021fsce}  & 41.63 & 43.08 & 56.60 & 77.52      \\
        MSOCL~\cite{chen2022multiscale}  &  44.82  &  46.45   & 51.19   &  75.49          \\
        SAE-FSDet~\cite{liu2024few} & 57.96 & 59.40 & 71.02 & 85.08  \\ \hline
        FSCE + Ours &47.96 &51.68 & 60.09 & 79.70 \\
        MSOCL + Ours & 61.69 & 70.78 & 73.38 &  79.29\\
        SAE-FSDet + Ours & 59.83 & 61.76 & 72.56 & 85.54 \\
        \hline
    \end{tabular}
\end{table}
The table \ref{tab:nwpu_comparation} demonstrates that our method achieves improved performance across various baseline models.
Our method improves the mAP by an average of 1.56\% on SAE-FSDet~\cite{liu2024few}, with a maximum detection performance improvement of 2.36\%.  
Notably, as the number of samples increases, the improvement in the model performance gradually diminishes, due to the model approaching its capacity limit.

\subsection{Ablation Studies}
To better understand the effectiveness of each proposed module in our proposed method, we conduct qualitative and quantitative experiments on the DIOR~\cite{li2020object} dataset.

\subsubsection{Ablation on components}
To analyze the importance of each module we proposed, we perform ablation studies on different components over the DIOR~\cite{li2020object} dataset.
\begin{table}[htbp]
    \centering
    \caption{FSOD performance on the DIOR dataset with different modules under the 3, 5, 10 and 20-shot setting}
    \label{tab:ablation_components}
    \renewcommand{\arraystretch}{1.3} 
    \scalebox{0.9}{
    \begin{tabular}{c|cc|cccc}
        \hline
        \multirow{2}{*}{Method} & \multicolumn{2}{|c|}{Components} & \multirow{2}{*}{3-shot} & \multirow{2}{*}{5-shot} & \multirow{2}{*}{10-shot} & \multirow{2}{*}{20-shot} \\ 
        \cline{2-3}
         & HIIM & Contrastive loss & & & & \\
        \hline
        Baseline & & &15.89 & 17.04 & 30.20 & 37.34\\
        Baseline$^\dag$ & & & 16.54 & 19.48 & 31.78& 39.45\\
        Ours &\checkmark &  & 21.56 & 26.97 & 35.78 & 39.67\\
        Ours &\checkmark &\checkmark & 22.09 & 29.85 & 36.66 & 40.81 \\
        \hline
    \end{tabular}}
\end{table}
We first adopt SAE-FSDet~\cite{liu2024few} as our baseline and report the detection results of it on the DIOR~\cite{li2020object} dataset in the first row. 
We then use text inversion to generate some instances, and conduct experiments by concatenating them with the original dataset as a new dataset. The experimental results on the new dataset are shown in the second row of Table \ref{tab:ablation_components}.
The experiments demonstrate that while instances generated through few-shot finetune-based diffusion can enhance FSOD performance, relying solely on CLIP features is insufficient for generating high-quality instances to help improve the performance of FSOD.
Next, we employ our DIG-FSOD with only HIIM to generate target instances that match the sample quantity, conduct experiments by concatenating them with the original dataset to form a new dataset and the experimental results are shown in the third row of Table \ref{tab:ablation_components}.
Finally, we incorporate an additional class contrastive loss to enhance the semantic consistency of the generated images.
The experimental results demonstrate that HIIM can significantly improve FSOD performance, while the class contrastive learning loss can further enhance the performance.	
\subsubsection{Ablation on HIIM}
HIIM aims to invert remote-sensing instances to conditional latent spaces. And to verify the effectiveness of different image encoder, we conduct ablation studies on DIOR dataset.\par{}
\begin{table}[htbp]
    \centering
    \caption{Ablation study for Hybrid Image Inversion Module design}
    \label{tab:ablation_on_feature_extractor}
    \renewcommand{\arraystretch}{1.3} 
    \scalebox{0.92}{
    \begin{tabular}{c|cc|cccc}
        \hline
        Method & $Enc_{global}$ & $Enc_{local}$   & 3-shot & 5-shot & 10-shot & 20-shot \\ 
        \hline
        Baseline & \ding{55} & \ding{55} & 15.89 & 17.04 & 30.20 & 37.34 \\
        Ours & CLIP & \ding{55} & 16.54 & 19.48 & 31.78 & 39.45\\
        Ours    & CLIP & CLIP & 20.99 & 28.25 & 34.26 & 39.78 \\
        Ours    & CLIP & DINOv2 & 22.09 & 29.85 & 36.66 & 40.81\\
        \hline
    \end{tabular}}
\end{table}
The ablation experiment results regarding the selection strategies for the global and local feature extractors are shown in Tab.~\ref{tab:ablation_on_feature_extractor}. 
The results in the first row present the performance of the baseline method, while the second line reports the results obtained with the data augmentation method and the results demonstrate that our data augmentation method can effectively improve detection performance on datasets with limited samples.
We subsequently integrated a local feature encoder to further enhance the detailed information in the generated images, as demonstrated in the third row of Tab.~\ref{tab:ablation_on_feature_extractor}.
Finally, we substitute the local feature encoder with the more fine-grained DINOv2~\cite{oquab2023dinov2}, as illustrated in the fourth row of Tab.~\ref{tab:ablation_on_feature_extractor}.
Compared to the baseline, our method achieves an average improvement of 7.23\%.	\par{}
\section{Conclusion}
This letter proposes a novel framework named DIG-FSOD, which aims to leverage a diffusion model to generate diverse remote sensing instances, to further improve the performance of few-shot object detectors. 
We employ a novel HIIM to simultaneously leverage both global semantic features and local detail features from the conditional images to synthesize higher-quality images. 
Furthermore, we incorporate a novel contrastive learning loss to enhance the consistency between the conditional image and its class.
Experiments on DIOR~\cite{li2020object} and NWPU VHR-10~\cite{cheng2014multi} datasets have validated the effectiveness of our proposed method.

\bibliographystyle{IEEEtran}
\bibliography{ref}

\end{document}